\documentclass{elsart}

\usepackage{graphicx}

\journal{J. Magn. Magn. Mater.}

\begin{document}

\begin{frontmatter}

\title{Electronic and magnetic  properties of the (111) surfaces of NiMnSb}

\author{I. Galanakis}
\ead{I.Galanakis@fz-juelich.de}

\address{Institut f\"ur Festk\"orperforschung, Forschungszentrum J\"ulich, D-52425 
J\"ulich, Germany}

\begin{abstract}
Using an ab-initio electronic structure method, I study the (111) surfaces 
of the half-metallic NiMnSb alloy. In all cases there is a very pronounced
surface state within the minority gap which destroys the half-metallicity 
This state survives for several atomic layers below the surface contrary to the 
(001) surfaces where surface states were  located only at the surface layer.  
The lower dimensionality of the surface leads in general to large enhancements
of the surface spin moments.
\end{abstract}
\end{frontmatter}

\section{Introduction}
\label{sec1}

The discovery of giant magnetoresistance in 1988 \cite{GMR} 
gave birth to  a new field  in condensed matter, the magneto- or spin--electronics
\cite{Zutic2004}.
Contrary to the conventional electronics it is the spin of the electron and not its
charge which plays the central role. 
Key compounds to maximize the efficient of the devices based on spintronics
are the so-called half-metallic  materials which are ferromagnets 
where there is a band gap at the Fermi level ($E_F$) for the minority spin
band while the majority spin band is metallic. 

Attractive candidates for half-metallic materials are the 
half-Heusler alloys and it is mainly NiMnSb which has attracted most of 
the attention. NiMnSb was also the first material to be predicted to be a half-metal
 in 1983 by de Groot and his collaborators \cite{groot}. 
There exist several other ab-initio calculations
on NiMnSb reproducing the results of de Groot \cite{bulkcalc} and Galanakis
\textit{et al.} showed that the gap arises from the hybridization between the $d$ 
orbitals of the Ni and Mn atoms \cite{iosifHalf}.
Its half-metallicity seems to be well-established in the case of single crystals;
infrared absorption \cite{Kirillova95}
and spin-polarized positron-annihilation \cite{Hanssen90}
gave evidence supporting it. 

The interest in the case of NiMnSb films has been focused mainly to films 
grown along the [001] direction. Several experimental groups have grown such films
\cite{001exper} and they  were found not be half-metallic 
\cite{001polar}; a maximum
value of 58\% for the spin-polarization of NiMnSb was obtained by
Soulen \textit{et al.} \cite{Soulen98}. Also several ab-initio calculations 
exist on the (001) surfaces films and all the results agree that the half-metallicity is lost
due to surface states \cite{iosifSurf,001calc}. On the other hand van 
Roy and collaborators
have managed to grow (111) films of NiMnSb on top of GaAs(111) substrates \
\cite{Roy1,Roy2}. These films
were found to contain inclusions of MnSb and NiSb \cite{Roy1} and to show a large amount
of defects \cite{Roy2} but their magnetic properties have not been studied.

In this communication I study the (111) surfaces of the
half-metallic NiMnSb Heusler alloy. 
I take into account all possible six terminations and show that 
in all cases there is a minority surface states in the region of the  gap.
In section \ref{sec2} I discuss the structure of the surface and the 
computational details and in section \ref{sec3} I present and analyze my results.
Finally in section \ref{sec4} I summarize and conclude.

\section{Computational method and structure}
\label{sec2}

In the calculations I used the the screened
Korringa-Kohn-Rostoker (KKR) Green's function method
\cite{zeller95,Pap02} within the atomic sphere approximation (ASA)
in conjunction with the local spin-density approximation \cite{vosko}
for the exchange-correlation potential \cite{Kohn}. The ASA calculations 
take into account the full charge density. 
To simulate the surface I used a slab with 33 metal layers embedded in
half-infinite vacuum from each side.  This
slab thickness is enough so that the layers in the middle exhibit
bulk properties; they show a spin-down gap of the same width as in
the bulk and the same relative position of the Fermi level and
finally the magnetic moments differ less than 0.01$\mu_B$ from the
bulk values. Here I should mention that this number of layers is much 
larger than the one needed in the case of the (001) surfaces \cite{iosifSurf} due to
the very intense surface states pinned at the Fermi level (see results in next section).
I have also converged the  {\bf q}$_\parallel$-space grid, the number of energy 
points and
the tight binding cluster so that the properties of the surfaces do not change
(similar DOS  and spin moments).  
So I  have used a
two-dimensional 30$\times$30 {\bf q}$_\parallel$-space grid to
perform the integrations in the first surface Brillouin zone. To evaluate
the charge density one has to integrate the Green's function over an
energy contour in the complex energy plane; for this 42 energy
points were needed. 
A tight-binding cluster of 51 atoms was used in the calculation of
the screened KKR structure constants \cite{zeller97}.  
Finally for the wavefunctions I took angular momentum up to
 $\ell_{max}=3$ into account  and for the
charge density up to $\ell_{max}=6$. 

\begin{figure}
  \begin{center}
\includegraphics[scale=0.6]{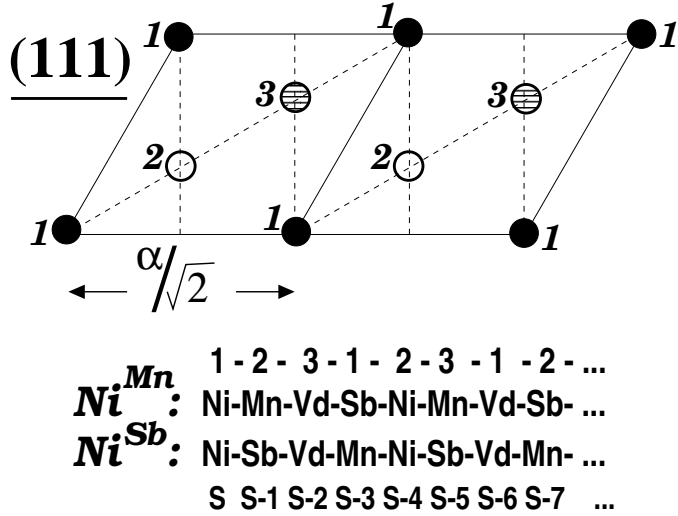}
  \end{center}
\caption{\label{fig1}
Schematic representations of the (111) surface. Each layers has a single atomic 
character. If I neglect the chemical character of the atoms in the $C1_b$ structure
of NiMnSb then the lattice is the   bcc. Note also that there are 
two different terminations for each element, \textit{e.g.} when the surface is 
terminated by Ni the subsurface layers can be either a Mn one or an Sb one. 
In the figure I show both of them together with the corresponding sequence of atoms.
``S'' stands for surface and ``1, 2, 3'' are the positions within the surface 
unit cell for succesive layers.}
\end{figure}

In figure \ref{fig1} I  present the structure of the
(111) surfaces in the case of NiMnSb. NiMnSb crystallizes in the $C1_b$ structure.
 If one neglects the difference in the chemical character of the atoms 
then the underlying lattice is a simple bcc. Thus along the [111] direction the 
alloy is consisted from alternating layers containing only one chemical element. 
In the case of a Ni terminated surface, as the one shown in the figure, there 
are two different possibilities: either to have a Mn subsurface layer 
or an Sb one.   I  denote the first case as Ni$^\mathrm{Mn}$ and the second 
as Ni$^\mathrm{Sb}$. Similarly Mn and Sb terminated surfaces can have either 
a Ni subsurface layer or one consisted from voids (denoted as Vd in the article).
Of course in real systems no voids exist but its use is obligatory in the calculations
to take correctly into account the charge distribution in the vacuum. The surface unit cell is a rhombus and when passing
from one layer to  a neighboring           one, the atoms slide 
along the big diagonal and are positioned at 1/3 or 2/3 of its length 
(in the figure atoms in successive layers are denoted by 1, 2 or 3).
I  have used in all my 
calculations the experimental lattice constant of NiMnSb of 5.909 \AA$\:$
\cite{Castel}. Finally I should mention that since my slab is made of 33 atoms
I have two inequivalent surfaces. For example if the one surface is the 
 Ni$^\mathrm{Mn}$ then the other one should be the  Ni$^\mathrm{Sb}$ surface.

\begin{table}
\caption{Atomic spin moments given in $\mu_B$ for the Ni terminated (111)
surfaces. ``S'' denotes the surface layer. 
 In parenthesis the bulk values.}\label{table1} 
\begin{tabular}{r|l|r|l|r} \hline
&  \multicolumn{2}{c|}{Ni$^\mathrm{Mn}$}&  \multicolumn{2}{c}{Ni$^\mathrm{Sb}$}
 \\  \hline
S &   Ni & 0.537 ( 0.268)& Ni &0.207  ( 0.268)\\
 S-1 &           Mn & 3.934 ( 3.700)& Sb &-0.060 (-0.069)\\
S-2 &          Vd & 0.058 ( 0.058)& Vd & 0.033 ( 0.058)\\
S-3&           Sb &-0.066 (-0.069) & Mn &3.709  ( 3.700) \\       \hline
S-4 & Ni & 0.252 ( 0.268)& Ni & 0.246 ( 0.268)\\
S-5&           Mn & 3.691 ( 3.700)& Sb & -0.074 (-0.069)\\
S-6&           Vd & 0.056 ( 0.058)& Vd & 0.058 ( 0.058)\\
S-7&           Sb & -0.062 (-0.069)& Mn & 3.682 ( 3.700)\\ \hline
\end{tabular}
\end{table}

\section{Results and discussion}
\label{sec3}

Surfaces can change the bulk properties severely, since the coordination of the
surface atoms is strongly reduced. Since half-metals can be considered as 
hybrids between the ferromagnetic metals and the semiconductors, 
two effects should be particularly relevant for their surfaces:
(i) for ferromagnets the moments of the surface atoms are strongly enhanced due to the
missing hybridization with the cut-off neighbors, and 
(ii) for semiconductors surface states
appear in the gap, such that the surface often becomes metallic. Also this is a
consequence of the reduced hybridization, leading to dangling bond states in the gap.
In Heusler alloys both phenomena occur simultaneously \cite{iosifSurf} but there are cases
like the Cr-terminated (001)   surfaces of the half-metallic zinc-blende CrAs or CrSe compounds,
where the large enhancement of the Cr spin moment kills the surface states \cite{iosifZB}.
     
\subsection{Ni terminated surfaces}
\label{sec3-1}

The first case which I will study are the Ni-terminated surfaces.
As it was already mentioned  there are two different possibilities: (i)
to have a Mn subsurface layer and then the sequence of the atoms
is Ni - Mn - Vd - Sb - Ni - Mn - ..., (ii) an Sb subsurface layer and then the
sequence is Ni - Sb - Vd - Mn - Ni - Sb - ...
The first one is denoted as Ni$^\mathrm{Mn}$  and the second one as 
 Ni$^\mathrm{Sb}$. In the bulk case Ni has four Mn and four Sb atoms 
as first neighbors. When I open the (111) surface the Ni atom at the 
surface layer loses four out of its 
eight first neighbors. In the case of Ni$^\mathrm{Mn}$ it loses three Sb atoms and
one Mn atom while in the Ni$^\mathrm{Sb}$ case one Sb and three Mn atoms. 

\begin{figure}
  \begin{center}
\includegraphics[scale=0.6]{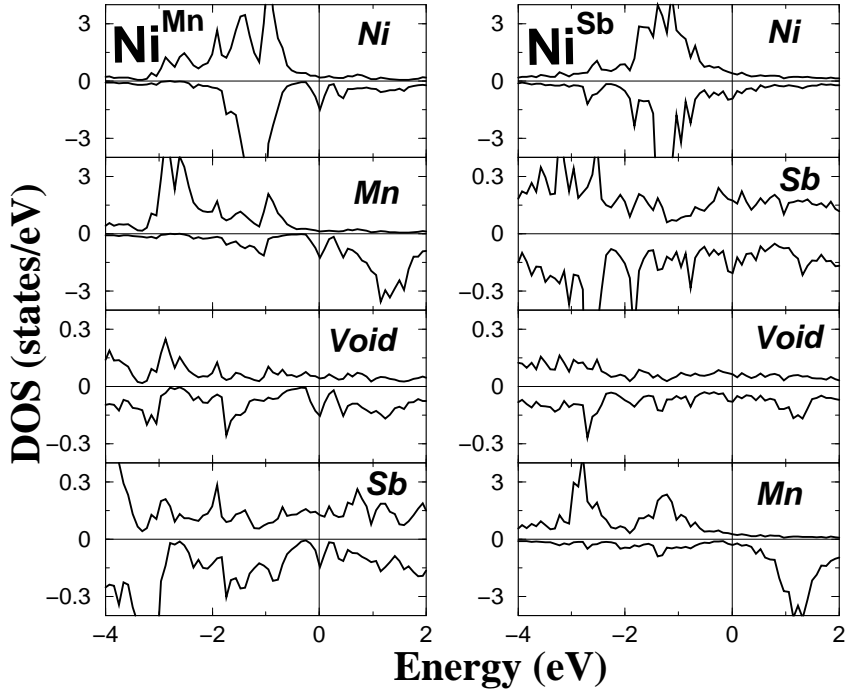}
  \end{center}
\caption{ \label{fig2}
Atom- and spin-resolved DOS for the two different surfaces terminating in Ni.
I show the DOS for the four upmost layers. The zero of the energy is chosen to
correspond to the Fermi level. Positive values of the DOS correspond to the majority spin
and negative to the minority.}
\end{figure}

In table \ref{table1} I have gathered the spin moments for the first eight layers 
and in parenthesis the spin moments for the bulk are given.
In the case of Ni$^\mathrm{Mn}$ both Ni and Mn atoms at the surface 
have very large moments with respect to both the bulk calculations and the 
 Ni$^\mathrm{Sb}$ case. Especially Ni moments is  doubled (0.54 $\mu_B$) 
with respect to the bulk value of 0.27$\mu_B$. In the case of the bulk NiMnSb
the minority gap is created by the hybridization between the $d$-orbitals of the 
Ni and Mn atoms, but the Sb atom plays also a crucial role since it provides states 
lower in energy than the $d$ bands which accommodate electrons of   the transition
metal atoms \cite{iosifHalf}. When I open the (111) surface terminated at  
Ni$^\mathrm{Mn}$ then each Ni surface atoms loses three out of the four Sb first neighbors
and they regain most of the charge accommodated in the  $p$ bands of Sb. These extra electrons
fill up mostly majority states increasing the Ni spin moment. But both majority
and minority Ni-bands are already completely occupied even in the case of bulk NiMnSb.
The only solution is that also Mn majority spin charge is increased to compensate for the
missing Sb neighbors although Mn and Sb atoms are second neighbors. This is 
helped by the fact that Mn and Ni majority $d$ states strongly hybridize forming
a common majority band as it was shown in reference \cite{iosifHalf}.
And thus the spin moment of Mn at the subsurface layer increases 
reaching the 3.93$\mu_B$ with respect to the bulk value of 3.70$\mu_B$. If one goes 
further away from the surface, the atoms have a bulklike environment and 
their spin moments are similar to the bulk case.
In the  Ni$^\mathrm{Sb}$ surface, Ni at the surface loses only one Sb first neighbor 
and the effect of the cut-off neighbors is much smaller. The moment is slightly smaller
than the bulk one mainly due to a surface state at the minority band shown in figure
\ref{fig2}. Already at the Sb subsurface atom one regains a bulklike behavior for 
the spin moment.

In figure \ref{fig2} I have gathered the spin-resolved density of states (DOS) 
for the four layers closest to the surface for both types of Ni termination. Note 
that I use a different grid for the DOS in the case of Mn and Ni atoms than for the 
Sb atom and the vacant site. For the  Ni$^\mathrm{Mn}$ termination, there is a minority surface 
state pinned exactly at the Fermi level which completely destroys the half-metallicity.
As mentioned above the population of the majority states increases and due to the 
exchange splitting the minority states are pushed higher in energy and this results 
to a very sharp shape of the surface state. This phenomenon is more pronounced
for the Mn atom at the subsurface layer, whose  occupied
minority states have a small weight,  and thus it presents a   much larger exchange splitting energy since this one 
scales with the spin magnetic moment. This surface state gradually weakens   and for 
the Ni atom at the S-4 position (not shown here) it practically vanishes. 

In the case
of the Ni$^\mathrm{Sb}$ surface the Ni spin moment is much smaller and the Mn atom is
deep in the substrate. Ni bands even move slightly higher in energy and thus 
the surface state is now much more extended  in the energy axis and cannot be well 
separated from the rest of the DOS. 
This situation is similar to the Ni-Void terminated (001) surface  studied 
in reference \cite{iosifSurf}.  There at the surface layer there are both Ni 
and void sites and the Ni surface DOS moves higher in energy with respect 
to the bulk killing the half-metallicity.

\subsection{Mn and Sb terminated surfaces}
\label{sec3-2}

In the second part of this section I will discuss the Mn and Sb terminated
 surfaces. As it was the case for the Ni terminated ones, there
are again two different possible terminations either having a Ni or
a void as the subsurface layer. In tables \ref{table2} and \ref{table3}
I have gathered the atomic spin moments in $\mu_B$ for all the possible cases
keeping the notation of table \ref{table1} to enumerate the different 
layers. 

\begin{table}
\caption{Same as table \ref{table1} for the Mn terminated surfaces.}\label{table2}  
\begin{tabular}{r|l|r|l|r} \hline
&  \multicolumn{2}{c|}{Mn$^\mathrm{Ni}$}&  \multicolumn{2}{c}{Mn$^\mathrm{Vd}$}
 \\  \hline
S &   Mn & 3.918 ( 3.700)& Mn & 4.202 ( 3.700)\\
S-1&           Ni &0.167  ( 0.268)& Vd &0.052 ( 0.058)\\
S-2 &          Sb &-0.103 (-0.069)& Sb &-0.067 (-0.069)\\
S-3&           Vd & 0.038 ( 0.058)& Ni & 0.260 ( 0.268)  \\       \hline
S-4 & Mn &3.614 ( 3.700) & Mn &3.679 ( 3.700)\\
S-5&           Ni &0.231 ( 0.268) & Vd &0.052 ( 0.058)\\
S-6&           Sb &-0.076 (-0.069)& Sb &-0.070 (-0.069)\\
S-7&           Vd &0.055  ( 0.058)& Ni &0.267 ( 0.268)\\ \hline
\end{tabular}
\end{table}

In the case of the Mn surfaces, Mn at the surface layer loses half of its
Sb second neighbors and similarly to what happened in the case 
of the Ni$^\mathrm{Mn}$ surface its spin moment is strongly enhanced
reaching the 3.92 $\mu_B$ for the Mn$^\mathrm{Ni}$ and the 4.20 
 $\mu_B$ for the Mn$^\mathrm{Vd}$ case. In the later case 
Mn has a Vd subsurface layer and thus the hybridization between
the Mn $d$-orbitals and the Sb $p$- and Ni $d$-orbitals is strongly reduced leading
to an increase of its spin moment with respect to the Mn$^\mathrm{Ni}$ case.
 The enhancement of the Mn spin moment at the interface is  also seen in the case of 
the (001) surfaces terminated in MnSb where the spin moment of the Mn atom at the
interface reaches a value of 4.02 $\mu_B$ \cite{iosifSurf}.
The atoms deeper in the surface quickly reach a bulklike behavior.

\begin{table}
\caption{Same as table \ref{table1} for the Sb terminated surfaces.} \label{table3}  
\begin{tabular}{r|l|r|l|r} \hline
&  \multicolumn{2}{c|}{Sb$^\mathrm{Ni}$}&  \multicolumn{2}{c}{Sb$^\mathrm{Vd}$}
 \\  \hline
S &   Sb & -0.134 (-0.069)& Sb &-0.198 (-0.069) \\
S-1&           Ni &  0.111 ( 0.268)& Vd & -0.005 ( 0.058)\\
S-2 &          Mn &  3.493 ( 3.700)& Mn & 3.606 ( 3.700)\\
S-3&           Vd & 0.038 ( 0.058)& Ni &  0.194 ( 0.268)\\       \hline
S-4& Sb & -0.067 (-0.069)& Sb & -0.080 (-0.069)\\
S-5&           Ni &  0.237 ( 0.268) & Vd &  0.051 ( 0.058)\\
S-6&           Mn & 3.704 ( 3.700)& Mn & 3.648 ( 3.700) \\
S-7&           Vd & 0.056 ( 0.058)& Ni &  0.260 ( 0.268)\\ \hline
\end{tabular}
\end{table}

Following the same arguments as for Mn, one can understand also the behavior  of the 
spin moments for the Sb terminated surfaces presented  in table \ref{table3}. 
The absolute value of the Sb spin moment at the surface layer increases with respect
to the bulk as was also the case for the MnSb terminated (001) surfaces 
\cite{iosifSurf}. When the subsurface layer is the Vd one, the hybridization effects
are less important and the Sb spin moment can reach a value of -0.20 $\mu_B$ almost
triple the bulk value of -0.07 $\mu_B$ and double the value for the (001) surface of
$-0.10$ $\mu_B$. The change in the Sb $p$-bands influences also through hybridization
the bands of the transition metal atoms for which now the minority bands population
increases leading to smaller spin moments of the Ni and Mn atoms at the subsurface 
layers. The phenomenon is more intense in the case of Sb$^\mathrm{Ni}$ where the Ni layer
is just below the Sb surface layer and the reduction in the spin moments of Ni and Sb is
much larger than the Sb$^\mathrm{Vd}$ case.

\begin{figure}
  \begin{center}
\includegraphics[scale=0.6]{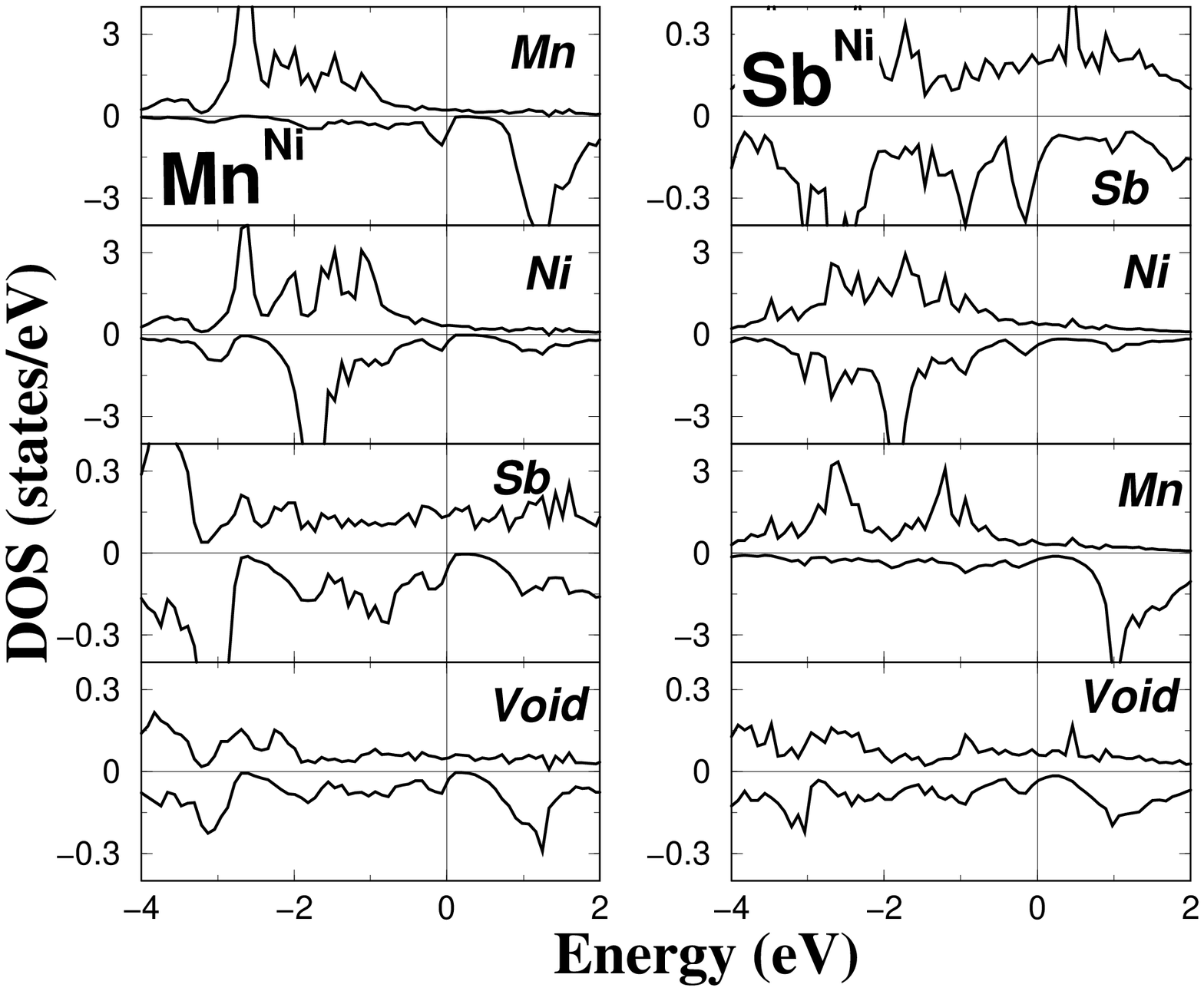}
  \end{center}
\caption{ \label{fig3}
Same as figure \ref{fig2} for the Mn$\mathrm{Ni}$ and Sb$\mathrm{Ni}$ 
terminated surfaces.}
\end{figure}

Finally in figure \ref{fig3} I have gathered the DOS for the first four layers
in the case of Ni subsurface layers. Results are similar also when I have the voids
as subsurface layer. In the case of the Mn terminated surface, there is a 
minority surface state pinned exactly at the Fermi level  which destroys the half-metallicity
and which survives also at the  Ni subsurface layer. But at the next Mn layer (not shown
here) it vanishes. Overall the DOS's are similar to the bulk case and the increase
of the Mn spin moment at the interface is taken care of high-energy lying 
majority antibonding $d$-states which in the bulk are above the Fermi level but now move 
below it pushing also somewhat lower in energy the 
majority bands \cite{iosifHalf}. Also in the case of the Sb terminated surface 
there is a minority surface state slightly below the Fermi level 
but which also destroys the half-metallicity at the surface. Its intensity is 
large also for the Ni at the subsurface layer but already for the Mn atom 
it starts to smear out.

\section{Summary and conclusions}
\label{sec4}

I have studied using an ab-initio technique the electronic and magnetic properties
of the (111) surface of NiMnSb taking into account different terminations.
In all cases there is a minority surface state which kills the half-metallicity 
at the surface. It is pinned at the Fermi level for the Ni and Mn terminated surfaces
but it is slightly below the Fermi level for the Sb one. It is localized close to
the surface region and normally vanishes within four atomic layers.

In the case of the Ni surface with Mn as subsurface layer, Ni$^\mathrm{Mn}$, 
the loss of three out of the
four Sb first neighbors leads to a doubling of the Ni spin moment while in the 
Ni$^\mathrm{Sb}$ case it is near the bulk value.  For the Mn and Sb 
terminations the lowering of the coordination increases the surface spin moments
and the enhancement is larger when there is a subsurface layer made by voids.

Finally I should add that from  point of view of transport a single surface state 
does not affect the magnetoconductance since the wavefunction is orthogonal to all 
bulk states incident to the surface. It is the interaction of the surface
state with other defect  states in the bulk and/or
with surface defects which makes these states conducting and leads to the low spinpolarization
values for films derived by Andreev reflection measurements.

\end{document}